 \documentclass[aps,pra,superscriptaddress,amsmath,amssymb,preprintnumbers,twocolumn,floatfix,showpacs,showkeys,10pt]{revtex4-1}
 \usepackage{amssymb} \usepackage{epsfig}
 \begin{document}
  \title{Non-Hermitian-assisted enhancement precision of parameter estimation }

\author{You-neng Guo}
\affiliation{ Department of Electronic and Communication Engineering, Changsha University, Changsha, Hunan
410022, People's Republic of China}
\affiliation{Hunan Province Key Laboratory of Applied Environmental Photocatalysis, Changsha University, Changsha, Hunan
410022, People's Republic of China}
\author{Mao-fa Fang}
\email{mffang@hunnu.cn}
\affiliation{ Key Laboratory of Low-Dimensional Quantum Structures and
Quantum Control of Ministry of Education, and Department of Physics,
Hunan Normal University, Changsha 410081, People's Republic of
China}
\author{Guo-you Wang}
\email{gywang04@163.com}
\affiliation{College of Science, Hunan University of Technology, Zhuzhou 412008, People's Republic of China}
\author{Jiang Hang}
\affiliation{Physical and photoelectric science department, Guangdong Ocean University, Zhanjiang, 524088, People's Republic of China}
\author{Ke Zeng}
\email{zk92@126.com}
\affiliation{ Department of Electronic and Communication Engineering, Changsha University, Changsha, Hunan
410022, People's Republic of China}
\affiliation{Hunan Province Key Laboratory of Applied Environmental Photocatalysis, Changsha University, Changsha, Hunan
410022, People's Republic of China}

\begin{abstract}
Recently, Zhong et al. [Phys. Rev. A 87, 022337(2013)] investigated the dynamics of quantum Fisher information (QFI) in the presence of decoherence. We here reform their results and propose two schemes to enhance and preserve the QFIs for a qubit system subjected to a decoherence noisy environment by applying $\textit{non-Hermitian}$ operator process. The proposed schemes are related to the $\textit{prior}$ (or the $\textit{post}$ ) non-Hermitian operator process corresponding to before (or after) the interest system suffers from decoherence, respectively. Resorting to the Bloch sphere representation, we derive the exact analytical expressions of the QFIs with respect to the amplitude parameter $\theta$ and the phase parameter $\phi$, and in detail investigate the influence of $\textit{non-Hermitian}$ operator parameters on the QFIs. Compared with pure decoherence process (without non-Hermitian operator process), we find that the $\textit{post non-Hermitian}$ operator process can potentially enhance and preserve the QFIs by choosing appropriate $\textit{non-Hermitian}$ operator parameters, while with the help of the $\textit{prior non-Hermitian}$ operator process one could completely eliminate the effect of decoherence to improve the parameters estimation. Finally, a generalized non-Hermitian operator parameters effect on the parameters estimation is also considered.
 \end{abstract}

 \maketitle
\section{Introduction}
Quantum Fisher information (QFI)\cite{Braunstein, Helstrom} which is one of the most important ingredients in parameter estimation fields has been extensively investigated in both the classical and quantum worlds. QFI characterizing the sensitivity of a state with respect to perturbation of the parameters is employed as a measure to quantify the precision of parameter estimation. The parameter estimation is the purpose of the information in a finite number of observation value to obtain highest precision of the parameter and design schemes to attain that precision which is bounded by the quantum Cram\'{e}r-Rao inequality,\cite{Helstrom,Holevo} i.e., $\Delta\phi\geq1/\sqrt{n \mathcal {F}_{\phi}}$, where $n$ is the number of experiments and $\mathcal {F}_{\phi}$ denotes the QFI of the parameter $\phi$.  This inequality indicates that the inverse of the QFI provides the lower bound of the error of the estimation.  On the other hand, QFI is formally generalized from the classical one, and there are several variants of quantum versions of Fisher
information\cite{Helstrom,Wigner,Fisher,Luo}, among which the one based on the symmetric logarithmic derivative operator has been used most widely.  For example, assume that a parameter $\phi$ is encoded in quantum states $\rho_{\phi}$,  and then the QFI of $\phi$ is defined as\cite{Helstrom,Holevo}
\begin{equation}
\label{e1}
\mathcal {F}_{\phi}=\textrm{Tr}(\rho_{\phi}\mathcal {L}_{\phi}^{2})=\textrm{Tr}[(\partial_{\phi} \rho_{\phi})L_{\phi}],
\end{equation}
where $\mathcal {L}_{\phi}$ is the so-called symmetric logarithmic
derivative, which is determined by
$\partial_{\phi}\rho_{\phi}=(\mathcal
{L}_{\phi}\rho_{\phi}+\rho_{\phi}\mathcal {L}_{\phi})/2$
with $\partial_{\phi}=\partial/\partial\phi$. Making use of the
spectrum decomposition, one can diagonalize the matrix as
$\rho_{\phi}=\Sigma_{n}\lambda_{n}|\psi_{n}\rangle\langle\psi_{n}|$.
And then the QFI can be rewritten as \cite{Liu0}
\begin{equation}
\label{e2}
\mathcal {F}_{\phi}=\sum_{n}\frac{(\partial_{\phi}\lambda_{n})^2}{\lambda_{n}}+\sum_{n}\lambda_{n}\mathcal{F}_{\phi,n}
-\sum_{n\neq m}\frac{8\lambda_{n}\lambda_{m}}{\lambda_n+\lambda_m}|\langle\psi_{n}|\partial_{\phi}\psi_{m}\rangle|^2,
\end{equation}
where $\mathcal {F}_{\phi,n}$ is the QFI for pure state $|\psi_{n}\rangle$ with the form
$\mathcal {F}_{\phi,n}=4[\langle\partial_{\phi}\psi_{n}|\partial_{\phi}\psi_{n}\rangle-|\langle\psi_{n}|\partial_{\phi}\psi_{n}\rangle|^2]$.
Given any single qubit states $\rho$, one can rewrite the form as $\rho = \frac{1}{2}(1+\vec{r}\cdot\hat{\sigma})$ in the Bloch sphere representation, where $\vec{r}=(r_{x}, r_{y}, r_{z})$ is the real Bloch vector
and $\hat{\sigma}=(\hat{\sigma}_{x}, \hat{\sigma}_{y},\hat{\sigma}_{z})$ denotes the Pauli matrices. According to Ref.\cite{Xiao}, a simple and explicit expression of QFI could be obtained in terms of the Bloch sphere representation $\mathcal{F}_{\phi}$
\begin{eqnarray}
\mathcal{F}_{\phi}=\left\{\begin{array}{cc}
|\partial_{\phi} \vec{r}|^{2}+\frac{(\vec{r}\cdot\partial_{\phi}\vec{r})^2}{1-|\vec{r}|^{2}}, & \mbox{ if } \, |\vec{r}|<1,\\
|\partial_{\phi}\vec{r}|^{2}, & \mbox{ if } \, |\vec{r}|=1.
\end{array}\right.
\label{eq0}
\end{eqnarray}

From the perspective of parameter estimation, the larger QFI represents the higher estimation precision in general. Hence, how to improve the QFI becomes a key problem to be solved. Making use of quantum resources such as coherence, entanglement and so on, one can find that QFI of quantum systems can provide much more sensitivity than the classical ones even to beat the shot noise limit in principle, in particular, under some circumstances, one can even achieve the Heisenberg limit~\cite{Boixo0,Napolitano1,Napolitano2}. However, in the realistic physical parameter estimation process, the quantum system unavoidably interacts with the surrounding environments, so that the QFIs dynamics decreases monotonically with time resulting in the decrease of parameter estimation precision.
In recent years, how to improve the estimation precision has become a significant issue in both experimentally and theoretically. Different protocols and strategies
have been proposed and realized to improve the precisions of various parameters.  For example, Tan et al. reported the improvement precision of parameter estimation in atom interferometer under dephasing noise by using dynamical decoupling pulses~\cite{Tan1,Tan2}. Zhang et al. proposed a scheme to enhance the precision of quantum estimation by increasing entanglement of the input state~\cite{Zhang}.  Inspired by the classical driving, Abdel-Khalek studied the QFI for a two-level atom system driven by a phase noise laser under non-Markovian dynamics~\cite{Abdel-Khalek}. Later, parameter estimation precision of a
non-Markovian dissipative two-state system by classical driving is also studied by Li~\cite{Li1}. Taking non-Markovian effect into consideration, a qubit system in non-Markovian environment becomes a good candidate for implementation of quantum optics schemes and information with high precision~\cite{Berrada1, Berrada2}. Applying quantum feedback, Zheng et al. analyzed the parameter precision of optimal quantum estimation of a dissipative qubit~\cite{Zheng}. Besides, the enhancement of the precision of phase estimation in quantum metrology is investigated by employing weak measurement and quantum measurement reversal~\cite{He, Xiao}.

So far, most previous literature studying the precision of parameter estimation have been mainly focused on the Hermitian or quasi-Hermitian systems which govern the interest system unitary evolution~\cite{Liu,Jing,Pang,Zhong}. In particular, others have been focused on estimating an overall multiplicative factor of a Hamiltonian~\cite{Lang,Jarzyna,Joo,ZhangYM}, for example, estimating $g$ in a Hamiltonian $gH$~\cite{Boixo,Roy}. In a recent work~\cite{Liu},  Liu et al.
studied a general Hamiltonian parameter estimation problem
where the QFI for an exponential form initial state involved in a Hermitian operator. They found that the QFI is totally determined by this Hermitian Hamiltonian and the initial state. However, to the best of our knowledge, few detailed investigations are concern on estimating a non-Hamiltonian parameter.
Motivated by these works, we here take advantage of the presence of $\textit{non-Hermitian}$ operator instead of Hermitian one. In particular, we would like to answer questions such as the following: For a $\textit{non-Hermitian}$ Hamiltonian, is it much more efficient than Hermitian one
to increase the precision of parameter estimation? How does the $\textit{non-Hermitian}$ process achieve better precision than Hermitian one?

To anticipate these questions, this paper extends parameter estimation to a $\textit{non-Hermitian}$ case, we propose two schemes to enhance the parameters estimation  by applying property of the non-Hermitian Hamiltonian process. The proposed schemes are related to apply the $\textit{prior}$ or the $\textit{post}$ non-Hermitian operators process. The former case corresponding to before the interest system suffers from decoherence channel, a non-Hermitian Hamiltonian operator is operated, and the later case is opposite.  In contrast with the results obtained in Ref.\cite{Zhong}, we derive the exact expressions of the QFIs with respect to the amplitude parameter $\theta$ and the phase parameter $\phi$ in terms of the Bloch sphere representation, and in detail investigate the influence of non-Hermitian operator on the QFIs. From our results, it turns out that (i)One can improve the precision of parameter estimation by applying the non-Hermitian operator process. (ii)The QFIs not only depend on the decoherence rate $\eta$, but also depend on the non-Hermitian parameter ($\alpha$, $\tau$). (iii)The $\textit{post}$ non-Hermitian operators process is superior to the $\textit{prior}$ one to enhance the precision of parameter estimation. Finally, a generalized non-Hermitian operator process effect on the parameters estimation is also considered.

\section*{Non-Hermitian operators processing}
\begin{figure}[htpb]
  \begin{center}
   \includegraphics[width=7.5cm]{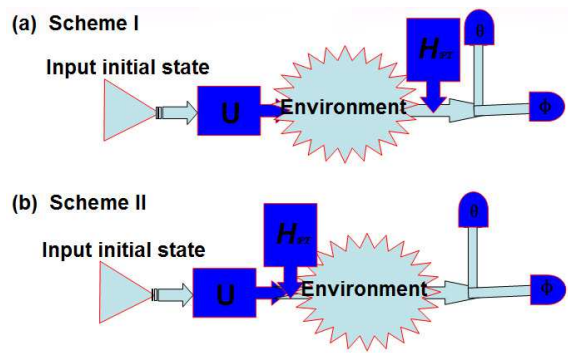}
   \caption{\label{fig1}Estimation of parameter in non-Hermitian operators processing. After the phase-gate operation, (a) a non-Hermitian operator is performed after the interest system interacts with a environment; (b) a non-Hermitian operator is applied before the interest system interacts
with a environment.}
\end{center}
\end{figure}
As we are all known, in conventional quantum mechanics, any physical obervables must be represented by self-adjoint or Hermitian operators in a Hilbert space, e.g. $\mathcal{H}_{}^{\dagger}=\mathcal{H}$. The Hamiltonian $\mathcal{H}$ of a closed system in conventional quantum mechanics is Hermitian, which requires not only the
eigenvalues are real, but also the time-evolution of the system is unitary. However, Bender et al. ~\cite{Bender}have recently found that the parity-time-symmetric Hamiltonian, which is non-Hermitian, can also have real energy spectra under some conditions.
Given a Hermitian Hamiltonian $\mathcal{H}$ of the interest system,
the evolution of the final state from time zero to $t$ is given by a map $|\psi(t)\rangle=U(t)|\psi(0)\rangle$, where $U(t)=\exp{(-i\mathcal{H})t}$ is a time evolution operator. According to conventional quantum theory, the evolution of the state is also generated by the Schr$\ddot{o}$dinger equation
\begin{equation}
i\frac{d}{dt}|\psi(t)\rangle=\mathcal{H}|\psi(t)\rangle ,  \label{master}
\end{equation}%
throughout this paper, we set $\hbar \equiv 1$. The corresponding density matrix of the entire system can be written as $\rho=|\psi(t)\rangle\langle\psi(t)|$, which satisfies $\rho=\rho_{}^{\dag}$ and $Tr\rho=1$.

In what follows we consider a different scenario by replacing the $\textit{Hermitian}$ Hamiltonian with a $\textit{non-Hermitian}$ one. For simplified, we assume this non-Hermitian Hamiltonian is~\cite{Bender}
\begin{equation}
\mathcal{H}=s%
\begin{pmatrix}
i\sin {\alpha } & 1 \\
1 & -i\sin {\alpha }%
\end{pmatrix},
\end{equation}%
where $s$ and $\alpha $ are real numbers. Based on Bender et al~\cite{Bender}, it is shown that
the non-Hermiticity of $\mathcal{H}$ can have real eigenvalues if it possesses parity-time symmetry, namely $[\mathcal{H}, \mathcal{PT}]$. Here the operator $\mathcal{P}$ is unitary and it is given by transform $x \rightarrow -x$, $p \rightarrow -p$, and the operator $\mathcal{T}$ is also unitary, one can make a transform $x \rightarrow x$, $p \rightarrow -p$, and $i \rightarrow -i$. Hence, it is not difficult to conclude the eigenvalues of $\mathcal{H}$ real satisfy the condition $|\alpha |\leq\pi /2$ and its eigenvalues are $E_{\pm}=\pm s\cos\alpha$. According to non-Hermitian quantum theory, to obtain the evolution of the system, one
should introduce a time-evolving operator $U_{}(t)= e^{-i\mathcal{H}t}$,
\begin{equation}
U_{}(t)=\frac{1}{\cos \alpha }%
\begin{pmatrix}
\cos (\tau-\alpha ) & -i\sin \tau \\
-i\sin \tau & \cos (\tau+\alpha )%
\end{pmatrix}%
,  \label{unitary}
\end{equation}%
Defining $\tau=s t\cos \alpha$ which is related to $s$ when fixed $t\cos \alpha$. Different from the conventional quantum theory, the evolution of the system with a non-Hermitian Hamiltonian is not trace-preserving in general
\begin{equation}
i\frac{\partial }{\partial t}\text{tr}(\rho )=2\text{tr}%
\left( \rho H\right).
\end{equation}%
Therefore, make sure the probability of normalization, we renormalize the system density operator
\begin{equation}
\Tilde{\rho}(t)=\frac{ U_{}(t) \rho
U_{}(t) ^{\dagger }}{\text{Tr}[\left(  U_{}(t) \rho
U_{}(t) ^{\dagger }\right]}.
\label{renorm22}
\end{equation}%
In the following, we utilize this non-Hermitian Hamiltonian process given by Eq.~(\ref{unitary}) to enhance and preserve QFI suffering from decoherence.

\subsection*{Enhancing precision of parameter estimation in non-Hermitian operators process}
\label{sec:3}
Before studying the precision of parameter estimation in non-Hermitian operator process, we firstly review the framework of parameter estimation. Usually the precision of parameter estimation is composed of three stages: the first stage is the preparation of initial state, the second stage is the sampling stage involved in the parameter code, and the third stage is the measurement corresponding to the parameter estimate. In following, we will adopt this framework to investigate the effect of non-Hermitian operators process on the parameter estimation of the amplitude parameter $\theta$ and the phase parameter $\phi$ in terms of QFIs for a two-level qubit subjected to a decoherence noisy environment.

\subsection*{Scheme $I$}

We start with the scheme $I$ as shown in Figure 1a. For simplicity, we consider the same scenario as Ref.~\cite{Zhong}, where an open system consisted of a two-level qubit subjected to an amplitude damping decoherence noisy environment after the phase gate $U_{\phi }=\left\vert g\right\rangle\left\langle g\right\vert +e^{i\phi }\left\vert e\right\rangle \left\langle e\right\vert$ is performed. We assume that the input state is a two-level superposition state $\left\vert \varphi \right\rangle=\cos\frac{\theta}{2}\left\vert g\right\rangle+\sin\frac{\theta}{2}\left\vert e\right\rangle$. After the phase gate operation $U_{\phi }$ operated on the input state $|\varphi\rangle $, the output state is given by $\rho =U_{\phi }\left\vert \varphi \right\rangle \left\langle \varphi \right\vert U_{\phi }^{\dagger }$ which contains two parameters, one is the amplitude parameter $\theta$  and the other is the phase parameter $\phi$. Next, let the state $\rho$ subject to the noisy environment,e.g. an amplitude damping decoherence environment, which is described as the quantum operation $\mathcal{E}$. Due to noisy decoherence, the state $\rho $ inevitably evolves into a mixed, namely $\rho_{}^{'}=\mathcal{E}(\rho)=\sum_{i=1,2}^{}E_{i}\rho E_{i}^{\dagger}$, where $E_{1}=\left\vert g\right\rangle\left\langle g\right\vert +\sqrt{1-\eta^2}\left\vert e\right\rangle \left\langle e\right\vert$, and $E_{2}=\eta\left\vert g\right\rangle\left\langle e\right\vert$.

In the next step of the scheme we perform the non-Hermiticity operation Eq.~(\ref{unitary}) on the open system after the implementation of the amplitude damping decoherence process. Then the finial evolution state $\rho_{out}$ of the system is obtained
\begin{equation}
\rho_{out}= U_{}(t) \mathcal{E}(\rho)
U_{}(t) ^{\dagger },
\label{renorm2}
\end{equation}%
As discussed above, the evolution of a system with a non-Hermitian Hamiltonian is
not trace-preserving and has to be re-normalized $\Tilde{\rho}_{I}(t)=(\rho_{out}/Tr\rho_{out})$. In the standard qubit basis $\{|g\rangle, |e\rangle\}$,
the non-zero elements of $\Tilde{\rho}_{I}(t)$ are described in terms of the Bloch sphere representation given by
\begin{widetext}
\begin{eqnarray}
\label{e30}
r_{x} &=& \frac{2\Delta\cos^2\alpha\cos\phi\sin\theta}{2+\Gamma\sin2\tau\sin2\alpha-2\sin\alpha(\cos2\tau\sin\alpha+2\Delta\sin^2\tau\sin\theta\sin\phi)},
\end{eqnarray}

\begin{eqnarray}
\label{e31}
r_{y} &=& \frac{\sin^2\tau(2\Delta\sin\theta\sin\phi-4\sin\alpha)+\Delta\sin\theta\sin\phi(1-2\cos^2\tau+\sin^2\alpha)-2\Gamma\cos\alpha\sin2\tau-\Delta\cos^2\alpha\sin\theta\sin\phi}
{2+\Gamma\sin2\tau\sin2\alpha-2\sin\alpha(\cos2\tau\sin\alpha+2\Delta\sin^2\tau\sin\theta\sin\phi)},
\end{eqnarray}

\begin{eqnarray}
\label{e32}
r_{z}&=&\frac{2\Gamma\cos2\tau\cos^2\alpha+\sin2\tau(\sin2\alpha-2\Delta\cos\alpha\sin\theta\sin\phi)}{2+\Gamma\sin2\tau\sin2\alpha-2\sin\alpha(\cos2\tau\sin\alpha+2\Delta\sin^2\tau\sin\theta\sin\phi)},
\end{eqnarray}
\end{widetext}
where $\Delta\equiv\sqrt{1-\eta^2}$ and $\Gamma\equiv1+\Delta^2(\cos\theta-1)$.
Finally, the scheme ends with one implementing a measurement on this finial state.
Substituting Eqs.(10)-(12) into Eq.~(\ref{eq0}), it is not difficult to calculate the QFIs of $\Tilde{\rho}_{I}(t)$ with respect to $\theta$ and $\phi$.
The analytical expressions of the QFIs can exactly be obtained
\begin{widetext}
\begin{eqnarray}
\mathcal{F}_{\theta}^{} &=& \frac{4\cos^2\alpha}{N}\{\Delta^2\cos^2\alpha\cos^2\phi[\cos\theta(2-2\cos2\tau\sin^2\alpha)+(\Delta^2+\eta^2\cos\theta)\sin2\tau\sin2\alpha]^2\nonumber\\
&+&4\cos^4\alpha[\Delta\cos2\tau\cos\alpha\cos\theta\sin\phi+\Delta\sin2\tau(\Delta^2\sin\alpha\sin\phi+\eta^2\cos\theta\sin\alpha\sin\phi-\Delta\sin\theta)]^2\nonumber\\
&+&4\eta^2\Delta^2\cos^2\alpha\sin^2\frac{\theta}{2}[(2-\cos2\tau+\cos2(\tau-\alpha))\cos\frac{\theta}{2}-4\Delta\sin^2\tau\sin\alpha\sin\phi\sin\frac{\theta}{2}]^2\nonumber\\
&+&[\Delta\sin\phi\sin2\alpha(\cos^2\tau+\eta^2\cos^2\tau\cos\theta+\eta^2+\eta^2\sin^2\tau)-2\Delta\sin\phi\sin2\tau\cos^2\alpha\cos\theta\nonumber\\
&+&\Delta^2\cos\alpha(1-2\cos2\tau-\cos2\alpha)\sin\theta+\Delta\sin2\alpha\sin\phi(\Delta^2\cos^2\tau-\eta^2\cos\theta(\sin^2\tau+1)-2)]^2\},   \label{master1}
\end{eqnarray}%
\begin{eqnarray}
\mathcal{F}_{\phi}^{}&=& \frac{4\cos^4\alpha}{N}\{256\eta^2\Delta^2\cos^2\frac{\theta}{2}\cos^2\phi\sin^4\tau\sin^2\alpha\sin^6\frac{\theta}{2}
+16\Delta^2\cos^2\phi\sin^2\tau\sin^2\theta(\cos\tau\cos\alpha+\Gamma\sin\tau\sin\alpha)^2\nonumber\\
&+&4\Delta^2\cos^2\alpha\cos^2\phi\sin^2\theta(\cos2\tau\cos\alpha+\Gamma\sin2\tau\sin\alpha)^2
+\sin^2\theta[\Delta(2-2\cos2\tau\sin^2\alpha+\Gamma\sin2\tau\sin2\alpha\sin\phi)\nonumber\\
&-&4\Delta^2\sin^2\tau\sin\alpha\sin\theta]^2\},  \label{master2}
\end{eqnarray}%
\end{widetext}
where $ N=[2-2\sin\alpha(2\Delta\sin^2\tau\sin\theta\sin\phi+\cos2\tau\sin\alpha)+\Gamma\sin2\tau\sin2\alpha]^4 $.
The calculations show the QFIs are deeply related to the input initial state parameters ($\theta$, $\phi$), the non-Hermitian parameters ($\alpha$, $\tau$) and the decoherence rate $\eta$.

Before investigating the non-Hermitian-assisted enhancement precision of parameter estimation, one should maximize the QFIs for all input states. It is easy to determine from Eqs.~(\ref{master1}) and ~(\ref{master2}) that the optimal input state is $\theta=\pi/2$ which is defined by the analytic calculation of $\frac{\partial\mathcal{F}}{\partial \theta}|_{\theta=\pi/2}=0$ and $\frac{\partial^2\mathcal{F}}{\partial^{2} \theta}|_{\theta=\pi/2}<0$.
Fig.2 shows the QFI with respect to parameter $\theta$ as a function of $\tau$ ($\tau=s t\cos\alpha$) for different phase parameter $\phi$. It can be seen QFI exhibits periodically oscillation. QFI possesses the maximum and minimum values depending on $\phi$. With the  phase parameter $\phi$ increasing, the transitions of QFI from concave into convex function. For $\phi=0$, QFI has the minimum values, while $\phi=\pi/2$, QFI has the maximum values. Hence, one might deduce that the optimal input state is $\theta=\pi/2$ and $\phi=\pi/2$. In fact, this deduction is true due to $\frac{\partial\mathcal{F}}{\partial \phi}|_{\phi=\pi/2}=0$ and $\frac{\partial^2\mathcal{F}}{\partial^{2} \phi}|_{\phi=\pi/2}<0$. However, it is worth noting that the initial value of QFI is independent of $\phi$, but dependent of the decoherence rate $\eta$. One can easily determine from Eqs.~(\ref{master1}) and ~(\ref{master2}) which reduces to $\mathcal{F}(\theta)=\mathcal{F}(\phi)=1-\eta^2$ at initial time $t=0$. The initial value of QFI varies  from 0 to 1. Moreover, for $\alpha=0$ or $\tau=0$($s=0$), Eqs.~(\ref{master1}) and ~(\ref{master2}) also reduce to $\mathcal{F}(\theta)=\mathcal{F}(\phi)=1-\eta^2$. These results are consistent with Ref~\cite{Zhong}, where the analytical expression of the QFI dynamics under amplitude damping decoherence channel without the $\textit{non-Hermitian}$ process. Note that the QFI with respect to parameter $\theta$ is the same as the one with respect to parameter $\phi$ when the optimal input state parameter is $\theta=\pi/2$. For simplicity, in the remainder of the section we only consider the QFI with respect to parameter $\theta$.

\begin{figure}[htpb]
  \begin{center}
   \includegraphics[width=5.5cm]{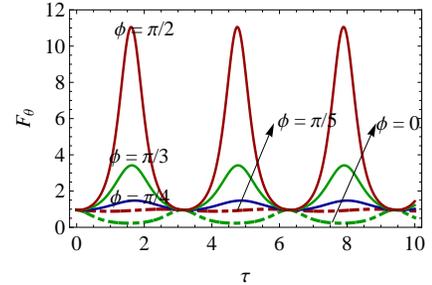}
   \caption{\label{fig2} QFI of the amplitude parameter $\theta$ as a function of $\tau$ ($\tau=s\cos\alpha t$) for different phase parameters $\phi$. Here we take $\theta=\pi/2$, $\alpha=\pi/5$ and $\eta=0.2$.}
\end{center}
\end{figure}

In following, we restrict our investigation to the
optimal input states ($\theta=\pi/2$ and $\phi=\pi/2$), and then Eq.~(\ref{master1}) reduces to
\begin{widetext}
\begin{eqnarray}
\label{eq3}
\mathcal{F}_{} &=& \frac{4\cos^4\alpha[\Delta(2-2\cos2\tau\sin^2\alpha+\eta^2\sin2\tau\sin2\alpha)-4\Delta^2\sin^2\tau\sin\alpha]^2}
{[2-2\sin\alpha(2\Delta\sin^2\tau+\cos2\tau\sin\alpha)+\eta^2\sin2\tau\sin2\alpha]^4}.
\end{eqnarray}
\end{widetext}
To quantify the efficiency of the $\textit{non-Hermitian}$ process on enhancing QFI, and concentrate only on how these process achieve better precision than Hermitian one, we introduce the difference of the QFI between with and without the implementation of $\textit{non-Hermitian}$ operator.
\begin{equation}
\Delta\mathcal{F}_{I}\equiv \mathcal{F}-\mathcal{F}_{}^{D}.
\label{eq4}
\end{equation}
where $\mathcal{F}_{}^{D}$ corresponds to the pure decoherence process. According to Ref.~\cite{Zhong}, the analytical expression of the QFI under amplitude damping decoherence channel is $\mathcal{F}_{}^{D}=1-\eta^2$. Based on Eq.~(\ref{eq4}), $\Delta\mathcal{F}>0$ implies enhancement precision of parameter estimation, while $\Delta\mathcal{F}<0$ corresponds to decrease precision of parameter estimation. Fig.3 clearly shows that QFIs with respect to the amplitude parameter $\theta$ (the phase parameter $\phi$) as a function of $\alpha$ and $\eta$ for $\tau=2.5$. From this graph, one can see clearly that the $\textit{post}$ non-Hermitian operator process can greatly enhance the QFI. If fixed parameter $\tau$, one can choose appropriate non-Hermitian operator parameters $\alpha$ to make $\Delta\mathcal{F}>0$, which reflects to enhance the parameters estimation. To illustrate parameters $\alpha$ effect more clearly, we plot QFI  as a function of $\tau$ for different parameters $\alpha$ in Fig.4. Several distinct features can be found from this graph: First, the evolution of QFI displays the same periodical oscillation independent of $\alpha$. Second, when $\alpha=0$, QFI is invariable and $\mathcal{F}_{I}=1-\eta^2$ for the fixed value $\eta$. Third, when $\alpha>0$, there exists $\mathcal{F}_{I}>1-\eta^2$, while $\alpha<0$ is related to $\mathcal{F}_{I}<1-\eta^2$.

In the above, only one of non-Hermitian parameters case is discussed. In Fig.5 we show QFIs as a function of $\tau$ and $\eta$ for $\alpha=\pi/5$. From this graph, it can be clearly seen that the $\textit{post}$ non-Hermitian operators process can greatly enhance the QFIs. By fixed parameter $\alpha$, one can choose appropriate  parameters $\tau$$(\tau=st\cos\alpha)$ to make $\Delta\mathcal{F}>0$. Besides, the optimal parameters $\alpha$ and $\tau$ could be obtained by calculating the following conditions:
$\partial\mathcal{F}_{I}/\partial \alpha=0$ and $\partial\mathcal{F}_{I}/\partial \tau=0$ correspond to $\partial^2\mathcal{F}_{I}/(\partial \alpha)^{2}<0$ and $\partial^2\mathcal{F}_{I}/(\partial \tau)^{2}<0$, respectively. By numerical calculation, we plot QFI with respect to the amplitude parameter $\theta$ (the phase parameter $\phi$) as a function of $\tau$ and $\alpha$ for $\eta=0.2$. It can be seen that the maximum $QFI$ corresponds to parameters $\alpha\approx \frac{5\pi}{16}$ and $\tau\approx 1.64$ as shown in Fig.6.
\begin{figure}[htpb]
  \begin{center}
   \includegraphics[width=5.5cm]{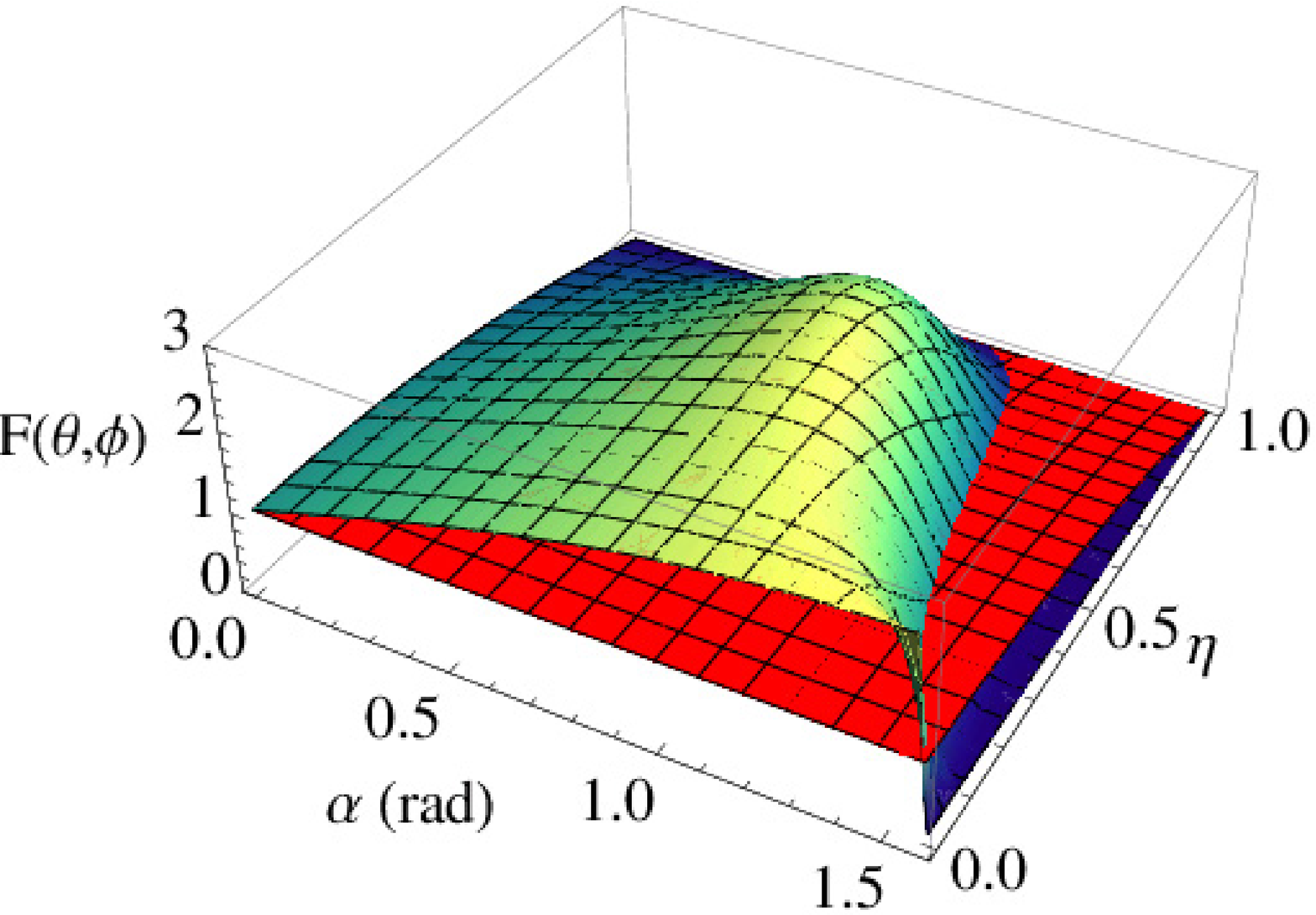}
   \caption{\label{fig3}QFIs of the amplitude parameter $\theta$ (the phase parameter $\phi$) as a function of $\alpha$ and $\eta$ with $\tau=2.5$. The red lower surface represents $\mathcal{F}_{}^{D}$ and the upper surface represents $\mathcal{F}_{I}$. Here we take $\theta=\pi/2$ and $\phi=\pi/2$.}
\end{center}
\end{figure}
\begin{figure}[htpb]
  \begin{center}
   \includegraphics[width=5.5cm]{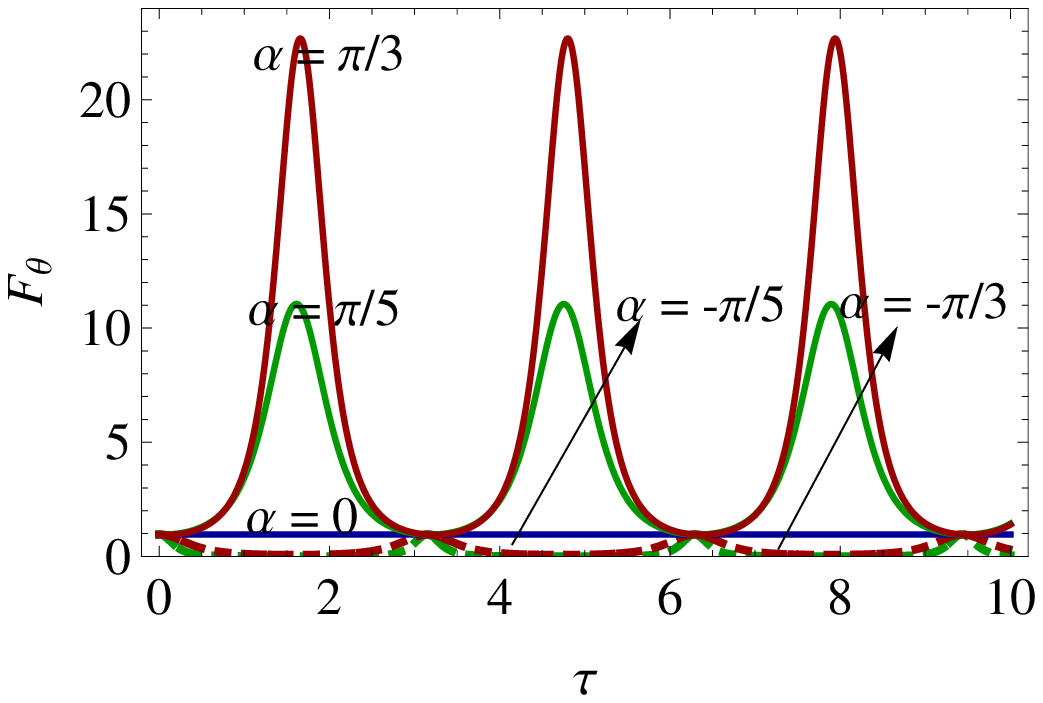}
   \caption{\label{fig4}QFI of the amplitude parameter $\theta$ as a function of $\tau$ for different parameter $\alpha$. Here we take $\theta=\pi/2$, $\phi=\pi/2$ and $\eta=0.2$.}
\end{center}
\end{figure}

\begin{figure}[htpb]
  \begin{center}
   \includegraphics[width=5.5cm]{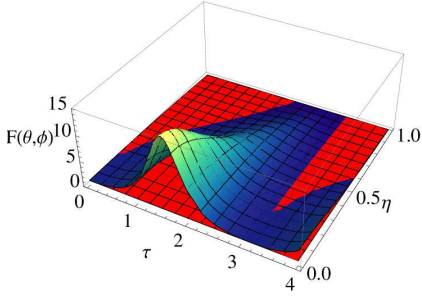}
   \caption{\label{fig5}QFIs of the amplitude parameter $\theta$ (the phase parameter $\phi$) as a function of $\tau$ and $\eta$ with $\alpha=\pi/5$. The red lower surface represents $\mathcal{F}_{}^{D}$ and the upper
surface represents $\mathcal{F}_{I}$. Here we take $\theta=\pi/2$ and $\phi=\pi/2$.}
\end{center}
\end{figure}

\begin{figure}[htpb]
  \begin{center}
   \includegraphics[width=5.5cm]{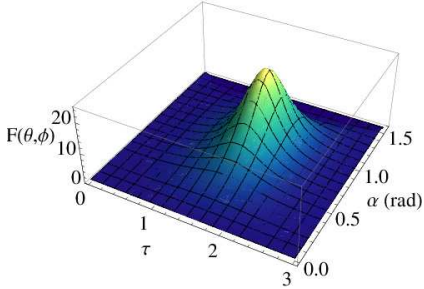}
   \caption{\label{fig6} QFIs of the amplitude parameter $\theta$ (the phase parameter $\phi$) as a function of $\tau$ and $\alpha$ with $\eta=0.2$. Here we take $\theta=\pi/2$ and $\phi=\pi/2$.}
\end{center}
\end{figure}

\subsection*{Scheme $II$}
Based on Scheme $I$ above, even though the precision of parameter estimation in $\textit{post non-Hermitian}$ operator process can be increased if the parameters $\alpha$ and $\tau$ are chosen appropriately, there are still some limitations. Following, we show an improved scheme that can completely circumvent the decoherence. The whole procedure is depicted in Fig.1 $(b)$, we first perform the $\mathcal{H}$ operator on the qubit system before it goes through the amplitude damping channel. The output state is given by $\Tilde{\rho}_{II}(t) =\frac{\sum_{i=1,2}E_{i}U(t)\rho U(t)_{}^{\dagger }E_{i }^{\dagger }}{Tr[\sum_{i=1,2}E_{i}U(t)\rho U(t)_{}^{\dagger }E_{i}^{\dagger }]}$, and the corresponding three Bloch vector components are
\begin{widetext}
\begin{eqnarray}
\label{e33}
r_{x} &=& \frac{\Delta\cos\phi\sin\theta}{\cos2\tau+\cos\theta\sin2\tau\tan\alpha+2\sec\alpha\sin^2\tau(\sec\alpha-\sin\theta\sin\phi\tan\alpha)},
\end{eqnarray}

\begin{eqnarray}
\label{e34}
r_{y} &=& -\frac{\Delta\sec^2\alpha[2\cos\alpha\cos\theta\sin2\tau+4\sin^2\tau\sin\alpha+(2\cos2\tau+\cos2\alpha-1)\sin\theta\sin\phi]}{2[\cos2\tau+\cos\theta\sin2\tau\tan\alpha+2\sec\alpha\sin^2\tau(\sec\alpha-\sin\theta\sin\phi\tan\alpha)]},
\end{eqnarray}

\begin{eqnarray}
\label{e35}
r_{z} &=&1-\frac{2\Delta^2\sec^2\alpha[\cos^2\frac{\theta}{2}\sin^2\tau+\cos^2(\tau+\alpha)\sin^2(\frac{\theta}{2})+\cos(\tau+\alpha)\sin\tau\sin\theta\sin\phi]}{\cos2\tau+\cos\theta\sin2\tau\tan\alpha+2\sec\alpha\sin^2\tau(\sec\alpha-\sin\theta\sin\phi\tan\alpha)}.
\end{eqnarray}
\end{widetext}
Similarly to the first scheme, we restrict our investigation to the
optimal input states ($\theta=\pi/2$ and $\phi=\pi/2$). Substituting  Eqs.(17)-(19) into Eq.~(\ref{eq0}), then the analytical
expressions of the QFI of parameter $\theta$ ($\phi$) can be obtained
\begin{eqnarray}
\label{e4}
\mathcal{F}_{II}=\frac{(1-\eta^2)(3-\cos2\alpha+4\sin\alpha)}{2(1+\cos2\tau\sin\alpha)^2}. \label{eq5}
\end{eqnarray}
From the above equation, it can be seen the QFI only depends on the non-Hermitian parameters ($\alpha$, $\tau$) and the decoherence rate $\eta$. In particular, when $\alpha=0$ or $\tau=0$, Eq.~(\ref{eq5}) reduces to $\mathcal{F}_{II}=\mathcal{F}_{}^{D}=1-\eta^2$.
In Fig.7 we plot QFIs with respect to the amplitude parameter $\theta$ (the phase parameter $\phi$) as a function of $\alpha$ and $\eta$ for $\tau=2.5$. One can see clearly that the $\textit{prior}$ non-Hermitian operator process could completely eliminate the effect of decoherence to improve the parameters estimation. The larger the non-Hermitian parameter $\alpha$ is, the higher the estimation precision.
On the other hand, we also plot QFIs with respect to the amplitude parameter $\theta$ (the phase parameter $\phi$) as a function of $\tau$ and $\eta$ for $\alpha=\pi/5$. From this graph, it is clearly shown that QFIs in the presence of the $\textit{prior}$ non-Hermitian operator process could be greatly improved.

In contrast to the results obtained in Ref.\cite{Zhong}, our schemes I and II all can enhance and preserve the QFIs. The underlying physical mechanism can be understood as follows: According to Eq.(7), the evolution of a system with a non-Hermitian Hamiltonian is not trace preserving. This means the non-Hermitian operation is not a genuine local operation in the conventional quantum world, so that the non-Hermitian natural has less than unity probability. Therefor, the enhancement of QFIs in our schemes can be attributed to the probabilistic nature of the non-Hermitian. From this viewpoint, the success probability of improving QFIs in scheme I is
\begin{eqnarray}
\label{e40}
\mathcal{P}_{I}&=&\sec\alpha[\sec\alpha+\eta^2\sin2\tau\sin\alpha\nonumber\\
&-&\tan\alpha(2\sqrt{1-\eta^2}\sin^2\tau+\cos2\tau\sin\alpha)],
\end{eqnarray}
which not only depends on the non-Hermitian parameter ($\alpha$, $\tau$), but also depends on the decoherence rate $\eta$. While The success probability of improving QFIs in scheme II is
\begin{eqnarray}
\label{e400}
\mathcal{P}_{II}=\cos2\tau+2\sec\alpha\sin^2\tau(\sec\alpha-\tan\alpha),
\end{eqnarray}
which is only dependent of the non-Hermitian parameters ($\alpha$, $\tau$) but independent of the decoherence rate $\eta$. However,
by compared with scheme I, scheme II is much better to enhance the parameters estimation. This indicates the input initial state performed the non-Hermitian operator is much more robust against the amplitude damping decoherence than the case where the input initial state in the absence of the non-Hermitian operator performing goes through the decoherent channel.
\begin{figure}[htpb]
  \begin{center}
   \includegraphics[width=5.5cm]{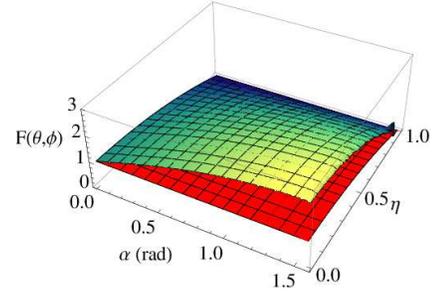}
   \caption{\label{fig7}Estimation of parameter in non-Hermitian operators process. QFI of the amplitude parameter $\theta$ (the phase parameter $\phi$) as a function of $\alpha$ and $\eta$ with $\tau=2.5$. The red lower surface represents QFI $\mathcal{F}_{}^{D}$ and the upper
surface represents QFI $\mathcal{F}$. Here we take $\theta=\pi/2$ and $\phi=\pi/2$.}
\end{center}
\end{figure}
\begin{figure}[htpb]
  \begin{center}
   \includegraphics[width=5.5cm]{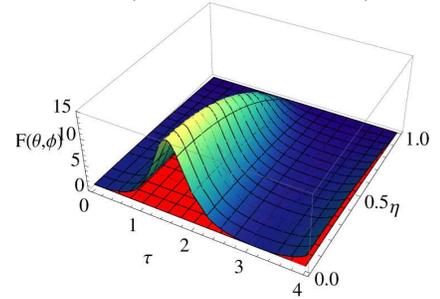}
   \caption{\label{fig8}Estimation of parameter in non-Hermitian operators process. QFI of the amplitude parameter $\theta$ (the phase parameter $\phi$) as a function of $\tau$ and $\eta$ with $\alpha=\pi/5$. The red lower surface represents QFI $\mathcal{F}_{}^{D}$ and the upper
surface represents QFI $\mathcal{F}$. Here we take $\theta=\pi/2$ and $\phi=\pi/2$.}
\end{center}
\end{figure}

\subsection*{Normal non-Hermitian Hamiltonian}
In the above discussion, we have assumed a specific non-Hermitian Hamiltonian with parity-time symmetry. Here, we consider a generalized non-Hermitian Hamiltonian
\begin{equation}
\mathcal{H'}=
\begin{pmatrix}
r e_{}^{i\alpha}+\delta & s \\
s & r e_{}^{-i\alpha}-\delta%
\end{pmatrix}%
,  \label{breakingPT}
\end{equation}%
where $r,s,\alpha,\delta$ are all real numbers. One can easily conclude this Hamiltonian violates parity-time symmetry and corresponds to the non-Hermitian operator process which is given
\begin{widetext}
\begin{equation}
U_{}^{'}(t)=e_{}^{-irt\cos\alpha}%
\begin{pmatrix}
e_{}^{-i\delta t}(\cos\omega t+\frac{2r\sin\alpha\sin\omega t}{\omega}) & \frac{2ie_{}^{-i\delta t}s\sin\omega t}{\omega} \\
-\frac{2ie_{}^{-i\delta t}s\sin\omega t}{\omega} & -e_{}^{-i\delta t}(\cos\omega t-\frac{2r\sin\alpha\sin\omega t}{\omega})%
\end{pmatrix}
\end{equation}%
\end{widetext}
where $\omega=\sqrt{s^2-r^2\sin^2\alpha}$. According to non-Hermitian quantum theory, the evolution of a system with a non-Hermitian Hamiltonian is
not trace-preserving. When we deal with the non-Hermitian process instead of Eq.(6), the renormalization procedure is also introduced. Therefore, if we repeat the processes of scheme I and II, the same phenomenon is also observed for a generalized non-Hermitian Hamiltonian. Namely, the precision of parameter estimation by applying the non-Hermitian operator process can be probabilistically improved if chosen appropriate non-Hermitian parameters $r,s,\alpha,\delta$.

\section*{Discussion}
In this work, we propose two schemes to enhance the parameters estimation by applying property of the non-Hermitian Hamiltonian procession.
The first one, where the actions of the $\textit{prior}$ non-Hermitian operator process before the system under decoherence, the second one dealing with the $\textit{post}$ non-Hermitian operators process after the system under decoherence. We calculate the exact expressions of the QFIs with respect to the amplitude parameter $\theta$ and the phase parameter $\phi$, and in detail investigate the influence of $\textit{non-Hermitian}$ operator parameters on the QFIs.
From our results, it turns out that: (i)The QFIs not only depend on the input initial state ($\theta$, $\phi$), but also depend on the non-Hermitian parameter ($\alpha$, $\tau$). (ii)One can improve the precision of parameter estimation by applying the $\textit{prior}$ non-Hermitian operator process. Finally, a generalized non-Hermitian operator process effect on the parameters estimation is also considered. Our results
provide an active way to suppress decoherence and enhance
the parameter-estimation precision, which is rather significant
in quantum precision measurement and quantum metrology.

\acknowledgments
This research is supported by the Funds of the National Natural
Science Foundation of China under (Grant No. 11374096), the Natural Science Foundation of Hunan Province (Grant No. 2016JJ2045), the Start-up Funds for Talent Introduction and Scientific Research of Changsha University 2015 (SF1504) and Scientific Research Project of Hunan Province Department of Education (16C0134 and 16C0469)

\label{app:eff-trans}

\end{document}